\documentstyle[twocolumn,aps,epsfig]{revtex}
\begin{document}
\sloppy

% INITIALIZE - DONT CHANGE
\title{Ordering phenomena and transport properties of Bi$_{1/2}$Sr$_{1/2}$MnO$_{3}$ single crystals}
\author{J. Hejtm\'anek, K. Kn\'{\i}\v{z}ek and Z. Jir\'ak}
\address{Institute of Physics of ASCR, Cukrovarnick\'a 10, 16253 Prague 6, Czech Republic}
\author{M. Hervieu and C. Martin}
\address{Laboratoire CRISMAT, UMR 6508, ISMRA, Bd du Mar\'echal Juin, 14050 Caen, France}
\author{M. Nev\v{r}iva and P. Beran}
\address{Institute of Chemical Technology, Technick\'a 5, 16628 Prague, Czech Republic}
\maketitle

\begin{abstract}
Measurements of the electrical and thermal conductivities, thermopower and paramagnetic
susceptibility have been performed on single crystal samples of Bi$_{1/2}$Sr$_{1/2}$MnO$_{3}$ and
complemented with the X-ray powder diffraction data. A pronounced hysteretic behavior, observed
below the cubic-to-orthorhombic transition at T$_{crit}=535$~K, is related to the onset of
long-range charge order at T$_{CO}=450$~K and its further evolution down to about 330~K. The
diffraction data suggest that the charge ordered state is formed by Zener pairs, represented by
Mn$^{4+}$ dimers linked by one extra $e_g$ electron, and is possibly stabilized by cooperative
Bi,Sr displacements. An extremely low thermal conductivity is observed down to the lowest
temperatures, without any recovery at the antiferromagnetic ordering temperature T$_N=150$~K. Such
behavior points to a presence of strong scatterers of phonons. Their possible origin can be linked
to "optical-like" oscillations which are associated with fluctuating charges within the Zener
pairs.
\end{abstract}

The interest in perovskite manganites with non-integral mean valence
(Mn$^{3+}$/Mn$^{4+}$) is associated mainly with their potential to become
ferromagnetically ordered and, at the same time, metallic. The origin of such
state may be seen in the so-called double-exchange (DE) mechanism
\cite{RefZener} that synergically favors fast migration of manganese $e_g$
electrons and ferromagnetic order due to a strong Hund intra-atomic coupling of
the itinerant $e_g$ electrons to the localized $t_{2g}^3$ electrons at Mn
sites. On the other hand, the tendency of DE to order the Mn spins
ferromagnetically via delocalization of $e_g$ electrons is opposed by strong
electron-lattice interactions which prefer the $e_g$ electron localization and
result in orbital ordering (Jahn-Teller effect of Mn$^{3+}$ ions), as well as
in the Mn$^{3+}$/Mn$^{4+}$ charge ordering effects for particular values of x
\cite{RefGoodenough,RefRadaelli1,RefJirak,RefRadaelli2,RefMizoguchi}.

The typical and often quoted examples of the charge ordered insulators are
perovskite Ln$_{1/2}$Ca$_{1/2}$MnO$_{3}$ (Ln~=~La or rare earths). The nature
of their state is still under debate. The generally accepted model is
Goodenough's orbital scheme \cite{RefGoodenough} based on a chess-board
arrangement of the Mn$^{3+}$ and Mn$^{4+}$ valences (Fig.~1a). An alternative
model, recently refined by Daoud-Aladine \cite{RefAziz1,RefAziz2}, is formed by
Zener pairs which are represented by Mn$^{4+}$ dimers linked by one extra $e_g$
electron (Fig.~1b). It should be noted that it is very difficult to distinguish
between these two structural models and their spin arrangements by the X-ray or
neutron powder diffraction methods, especially for systems with
discommensuration defects. In the single crystal diffractometry, the main
obstacle is the multidomain character of the charge ordered phase.
Nevertheless, for few compounds the conventional chess-board model was
unambiguously confirmed even if the actual extent of the charge
disproportionation is still uncertain (see e.g. \cite{RefJirak,RefDamay}). We
focus the present paper to a distinct case of Bi$_{1/2}$Sr$_{1/2}$MnO$_{3}$
which exhibits unusually high ordering temperature and we argue that its
ordered state can be described by the alternative model.

Previous work on the Bi$_{1/2}$Sr$_{1/2}$MnO$_{3}$ system reported charge ordering at
T$_{CO}=475$~K and pointed to the possible role of the Bi$^{3+}$ lone electron pair in the
transition \cite{RefGarcia}. The characteristic lattice deformation evolves with decreasing
temperature and saturates at about 300~K. The fully established charge order at room temperature
allowed an electron microscopy study to be performed at atomic resolution \cite{RefHervieu}. It was
shown that the structure of this manganite consists of double stripes of manganese octahedra, and
consequently differs from the conventional model. In a low-temperature study an antiferromagnetic
ordering of the CE type was detected below T$_N=155$~K while a minor part of the sample slightly
richer on Mn$^{4+}$ showed an A-type order \cite{RefFrontera}. In contrast to these studies,
performed on samples prepared by a ceramic route, the present experiments were undertaken on single
crystals grown from high-temperature solutions \cite{RefNevriva}. The electrical and thermal
conductivities and the thermoelectric power were measured between $20-900$~K using the four point
steady-state methods. The high temperature measurements were performed in air and the results were
registered both on warming and cooling. The structural characterization was done on pulverized
material using the Bruker~D8 X-ray diffractometer with CuK$\alpha$ radiation.

The chemical composition of the Bi$_{1/2}$Sr$_{1/2}$MnO$_{3}$ crystals was verified by Rietveld
refinements of powder diffraction data and confirmed (within the experimental error $\pm 0.02$) by
electron microanalysis and atomic absorption spectroscopy. In contrast to common $Pbnm$
orthoperovskites, a simpler tilt pattern of the $Ibmm$ symmetry was determined in the present
system. The determination of the charge ordered superstructure was attempted in the orthorhombic
cell doubled along the b-axis by using two variants - space group $Pnmm$ (conventional model) and
$Pnnm$ (alternative model). The latter model gave a better fit and more realistic oxygen
coordination of the Bi,Sr sites confirming that the ordered state in Bi$_{1/2}$Sr$_{1/2}$MnO$_{3}$
is indeed formed by Zener pairs as depicted in Fig.~1b.  The picture further shows that the
resulting superstructure is associated with a marked transversal modulation in which the
interconnected MnO$_{6}$ octahedra and the interpolated Bi,Sr are displaced in an opposite phase.
The amplitudes of the structural modulation are actually of about 0.1~\AA. These findings are in
distinction to the conventional charge ordering in other Ln$_{1/2}$A$_{1/2}$MnO$_{3}$ manganites
(A~=~Ca, Sr), in which similar modulations are in phase, corresponding to a homogeneous waving of
the perovskite lattice as a whole. The different behavior of Bi$_{1/2}$Sr$_{1/2}$MnO$_{3}$ thus
strongly suggests that the interpolated large cations, namely Bi$^{3+}$, play an active role in the
charge ordering and are at the root of the exceptionally high charge-ordering temperature T$_{CO}$.
A more detailed inspection of the structural results reveals two different Bi,Sr sites. One, shown
at y~=~0 in Fig.~1b, is slightly shifted upon the charge ordering towards equatorial oxygens of
neighboring MnO$_{6}$ octahedra along x. The second one, shown at y~=~0.5, makes an opposite
displacement and approaches thus the apical oxygen. Based on the refined Bi,Sr coordinates and
Debye-Waller factors it is possible to conclude that the shortest local Bi-O distances are about
2.4~\AA. Therefore, there remains an open question of the stereoactivity of the Bi$^{3+}$ lone
electron pair which is generally characterized with $2-5$ very short Bi-O bonds of 2.2~\AA\ only.

The resistivity data (Fig.~2) show that Bi$_{1/2}$Sr$_{1/2}$MnO$_{3}$ is essentially insulating at
low temperatures while at high temperatures well above T$_{CO}$ the electrical resistivity
decreases below 10~m$\Omega$cm. It is worth mentioning that the observed resistivity is 10$^3$
lower than the values reported earlier on ceramic samples \cite{RefFrontera}. The metal-insulator
transition is observed at T$_{crit}=535$~K, \textit{i.e.} above the reported charge ordering
temperature, concurrently with a change from the high-temperature cubic $Pm3m$ to the orthorhombic
$Ibmm$ structure (see the dependence of lattice parameters in Fig.~3). The transition is
accompanied by a hysteresis both in the resistivity and in the lattice properties. Closer
inspection shows that a marked hysteresis develops below 450~K, which can be related (in reference
to the previous work \cite{RefGarcia}) to the onset of the long-range charge order at T$_{CO}$ .
The data further indicate that this broad hysteretic region, characteristic for the 1-st order
transition, extends down to $\sim$330~K.

The character of the charge carrier conduction and the evolution of the phase transitions at
T$_{crit}=535$~K, T$_{CO}=450$~K and T$_{N}=150$~K are further evidenced by means of the
thermopower data, shown also in Fig.~2. Here, the M-I transition at T$_{crit}$  is accompanied by a
sharp decrease of the temperature independent thermopower (S~$\simeq -35 \mu$VK$^{-1}$), which
characterizes the adiabatic hopping conduction at higher temperatures. The large negative values of
the thermoelectric power below T$_{CO}$ suggest that the number of conducting electrons is
significantly decreased upon charge ordering. On further cooling below the N\'eel temperature
T$_{N}$, no substantial change of the electrical resistivity is observed. On the other hand, the
value of the thermopower undergoes a distinct decrease at T$_{N}$ before a final thermodynamic
turnover towards zero. This behavior indicates an additional charge carrier localization
conditioned by the long range antiferromagnetic order. The low temperature evolution of the
thermoelectric power between T$_{N}$ and 330~K likely mirrors the variable range hopping mechanism
of the charge carrier transport when the similar behavior was previously observed for
Sm$_{0.5}$Ca$_{0.5}$MnO$_{3}$ ceramics \cite{RefHejtmanek_SmCa}.

Finally, the temperature dependence of the thermal conductivity is displayed in Fig.~4. In the
charge ordered phase, the thermal transport is solely due to phonon contribution and its low
absolute value, which does not show any recovery below T$_N$, points to an extremely short mean
free path of the heat-carrying phonons. This becomes obvious namely in comparison with
substantially larger thermal conductivity observed in the conventional charge ordered system
Pr$_{0.5}$Ca$_{0.5}$MnO$_{3}$ or with even larger phononic conductivity observed in the
ferromagnetic single crystal Pr$_{0.52}$Sr$_{0.48}$MnO$_{3}$ (see Fig.~4).

In summary, the complex study of the Bi$_{1/2}$Sr$_{1/2}$MnO$_{3}$ crystals reveals three phase
transitions which manifest themselves in structural, transport and magnetic properties. These are:
the cubic-to-orthorhombic transition at T$_{crit}=535$~K, the onset of long-range charge order at
T$_{CO} \sim 450$~K and the antiferromagnetic ordering at T$_{N} = 150$~K. The structural study and
the behavior of inverse susceptibility (see caption of Fig.~2.) suggest that the charge ordered
state in Bi$_{1/2}$Sr$_{1/2}$MnO$_{3}$ is formed by Zener pairs and the resulting superstructure is
associated with a special kind of Bi,Sr displacements acting cooperatively with the $e_g$ electron
ordering. The low thermal conductivity in the ordered phase evidences a presence of strong
scatterers of phonons, persisting down to the lowest temperatures. Their possible origin can be
linked to "optical-like" oscillations which are associated with fluctuating charges within the
Zener pairs.

\acknowledgements This work was supported by grants A1010202 and A1010004 of
Grant Agency of Academy of Sciences of the Czech Republic.

\begin{figure}
 \epsfxsize=8.0cm \epsfbox{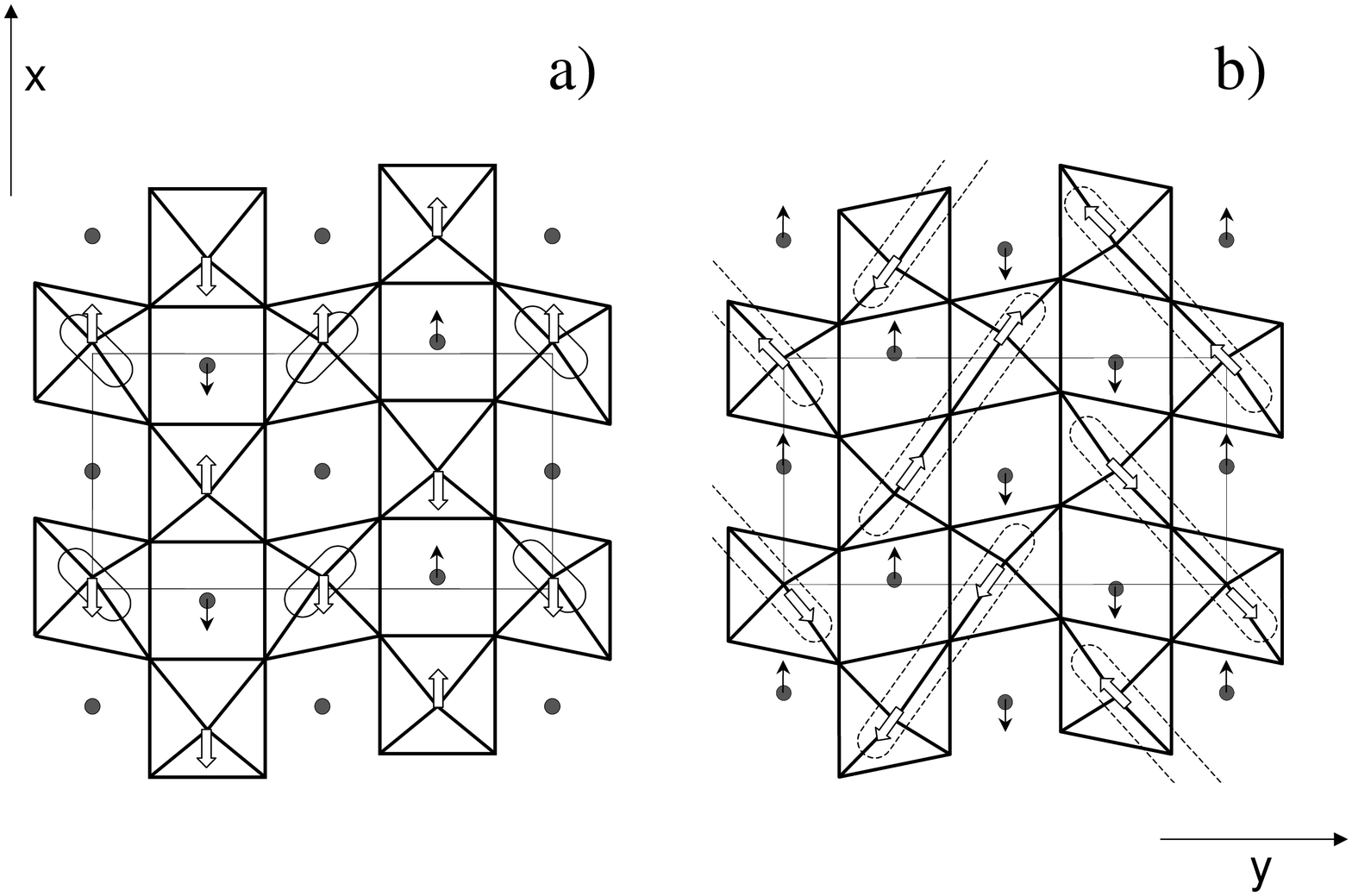}
\caption {The (001) perovskite layer in the charge ordered state - (a) conventional model realized
in Pr$_{0.5}$Sr$_{0.41}$Ca$_{0.09}$MnO$_{3}$ \protect\cite{RefDamay} and (b) model of Zener pairs
suggested for Bi$_{1/2}$Sr$_{1/2}$MnO$_{3}$. The $e_g$ electron densities in MnO$_{6}$ octahedra at
plane z=0 are schematically shown by the oval shapes; the main displacements of Bi,Sr sites at
z=1/4 are marked by arrows. The hollow arrows show the spin ordering - (a) antiferromagnetism of
the CE type and (b) non-colinear structure expected for the Zener-pair model. (In the plane z=1/2
the spins are reversed.) These magnetic arrangements are closely related, since case (b) can be
viewed as a superposition of two displaced CE arrangements with spins along x and y, respectively.}
 \label{fig1}
\end{figure}

\begin{figure}
 \epsfxsize=8.0cm \epsfbox{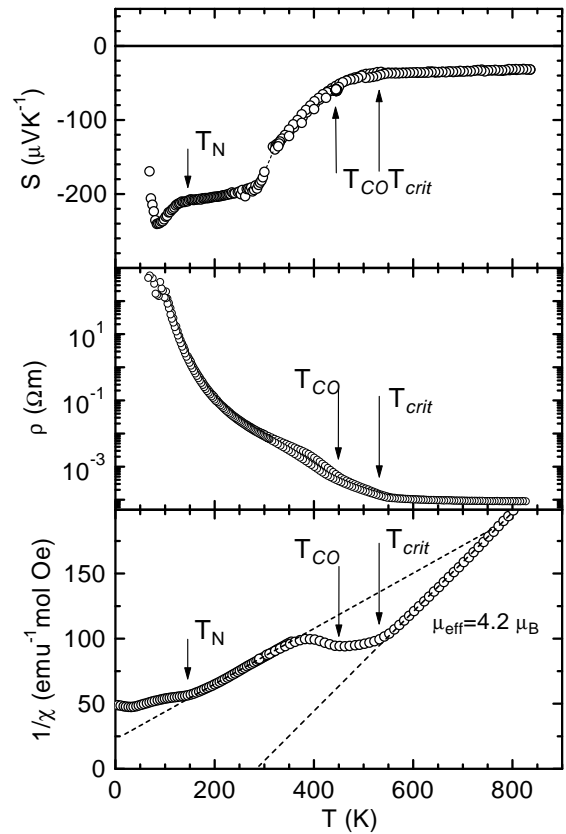}
\caption{The temperature dependence of the thermoelectric power (upper panel), electrical
resistivity (middle panel) and inverse magnetic susceptibility (lower panel) of the
Bi$_{1/2}$Sr$_{1/2}$MnO$_{3}$ crystal. The slope of $1/\chi$ in the charge ordered state for
$150-300$~K closely approaches the value for Zener pairs with total spin S~$=7/2$
($\mu_{exp}=7.83$~$\mu_B$, $\mu_{theor}=7.94$~$\mu_B$).}
 \label{fig2}
\end{figure}

\newpage

\begin{figure}
 \epsfxsize=8.0cm\epsfbox{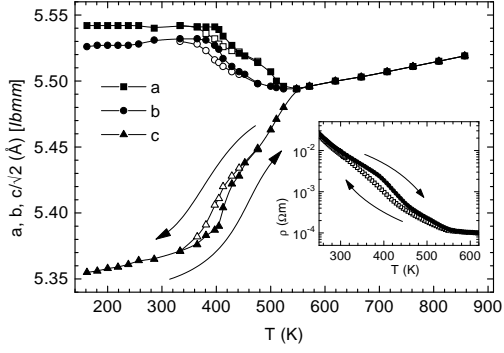}
\caption{The temperature dependence of lattice parameters in Bi$_{1/2}$Sr$_{1/2}$MnO$_{3}$. (The
inset shows a detail of the resistivity hysteresis upon the charge ordering.) }
 \label{fig3}
\end{figure}

\begin{figure}
 \epsfxsize=8.0cm \epsfbox{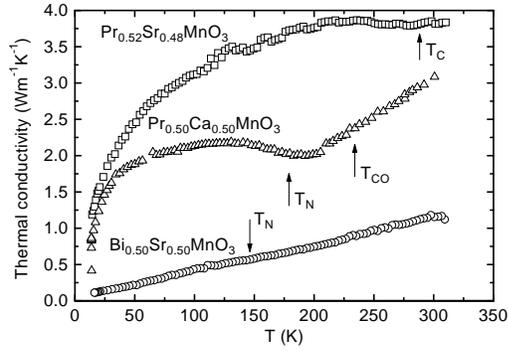}
\caption{The temperature dependence of the thermal conductivity in Bi$_{1/2}$Sr$_{1/2}$MnO$_{3}$
compared to the data on charge ordered antiferromagnet Pr$_{1/2}$Ca$_{1/2}$MnO$_{3}$
(T$_{CO}=240$~K, T$_{N}=180$~K) and ferromagnet Pr$_{0.52}$Sr$_{0.48}$MnO$_{3}$ (T$_{C}=295$~K;
electronic part of thermal conductivity is subtracted).}
 \label{fig4}
\end{figure}

\end{document}